%% file: hensler.jenam2010.tex
\documentclass[graybox]{svmult}

\usepackage{mathptmx}       
\usepackage{helvet}         
\usepackage{courier}        
\usepackage{type1cm}        
\usepackage{makeidx}         
\usepackage{graphicx}        
\usepackage{multicol}        
\usepackage[bottom]{footmisc}

\makeindex             

\begin{document}

\def\HI{H{\sc i} }
\def\HII{H{\sc ii} }
\def\Ha{{\rm H}\alpha }
\def\rg{\rho_{\rm g} }
\def\Msun{M_{\odot} }
\def\Mpc2{\Msun/pc^2}
\def\tff{\tau_{\rm ff} }
\def\tSF{\tau_{\rm SF} } 
\def\Zsun{Z$_{\odot}$ }
\def\OH{{12\-+log(O/H)} }
\def\lNO{log(N/O) }
\def\cd{{\it chemo-dynamical} }

\title*{Morphological Mutations of Dwarf Galaxies}
\author{Gerhard Hensler}
\institute{Gerhard Hensler \at University of Vienna, Institute of Astronomy, Austria, \email{gerhard.hensler@univie.ac.at}
}
%
%
\maketitle

\abstract{
Dwarf galaxies (DGs) are extremely challenging objects in 
extragalactic astrophysics. They are expected to originate as 
the first units in Cold Dark-Matter cosmology.
They are the galaxy type most sensitive to environmental influences 
and their division into multiple types with various properties 
have invoked the picture of their variant morphological
transformations. Detailed observations reveal characteristics
which allow to deduce the evolutionary paths and to witness how the
environment has affected the evolution. Here we review 
peculiarities of general morphological DG types and refer to
processes which can deplete gas-rich irregular DGs leading 
to dwarf ellipticals, while gas replenishment implies an 
evolutionary cycling.
Finally, as the less understood DG types the Milky Way satellite 
dwarf spheroidal galaxies are discussed in the context of 
transformation.}

\section{Introduction}
\label{intro}

By the continuous growth of telescope size and advanced
detector sensitivity the panchromatic view of galaxies is enabling 
us since the HST time to trace the evolution of massive 
galaxies observationally back to high redshifts. As examples the 
existence of intact gas-rich galaxy disks around redshift 2 has 
provided us a new insight into the gas accumulation and causes for 
the high star-formation rates (SFRs). Although dwarf galaxies (DGs)
also exist already at that early epoch, but because of their 
faintness, those observations are not as feasible for them so 
that our wisdom of DG formation and evolution depends on assumptions 
from numerical simulations and from their comparison with stellar 
population studies of DGs in the local universe. Nevertheless, due 
to the improved observational facilities also for the DGs, details of 
their properties have affected our picture of their formation 
and evolution. The first impression from decades ago, that DGs 
possess simple structures and evolve morphologically clearly 
separated, has changed totally in the sense that 
a classical morphological division of them is meaningless in the view 
of the variety of DG types: there are e.g. dwarf irregular galaxies
(dIrrs) with exceedingly strong star formation (SF), called 
starburst DGs (SBDGs), and also short but intense epochs of SF 
in the past, dwarf elliptical galaxies (dEs) with recent SF or 
central gas content, and last 
but not least, dwarf spheroidals (dSphs) at the faint end of dEs 
as satellite galaxies down to about -5$^m$. 
''Normal'' DGs have a brightness range between M$_V \ge -18^m$ 
to -10$^m$. 

For this brightness reason, dSphs are only detectible within the 
Local Group by refined search algorithms from surveys as e.g. SDSS. 
Also in the Virgo Cluster an archival work of detailed dE properties 
is expensive in observing time. In their studies of Virgo cluster 
DGs already Sandage \& Biggeli \cite{sb84} found that dEs dominate 
the cluster galaxy population by far, in contrast to their number
fraction in the field where dIrrs are the most common DGs. 
This fact cannot be interpreted
from the different local origins of DGs but because matter
accumulates to clusters also dIrrs fall in from the cosmic web
continuously whereby they have to change their morphology. 
Not only because of such morphological mutation but also due to 
the occurrence of enhanced SF in dIrrs, \cite{sb84} emphasized 
already the necessity of various links between the DG types by 
morphological transitions.

From the $\Lambda$CDM cosmology the baryonic matter should settle
within Dark Matter (DM) halos, which originally preferred to 
form low-mass subhalos and hierarchically accumulate to massive 
galaxies. 
If the baryonic matter would follow this bottom-up structure 
formation, the subhalos should also assemble their gas at first
and by this also evolve with SF to become the oldest galactic 
objects in the universe. 
That this picture seems to be too naive is simply understandable 
by three major physical principles: 

\noindent
1. The gas assembly timescale should behave as the free-fall 
timescale $\tff$, namely, dependent on the gas density as 
$\rg^{-1/2}$, because gas is accreted through gravitation. 
Whether this accretion leads to the same enhancement of SF in DGs 
as observed and theoretically expected \cite{kho09} is still a 
matter of debate (see also sect. 2).
Because the virial temperature in the DG gravitational potential
does not accomplish values above 10$^5$ K, on the one hand, 
cold accretion \cite{dek09} is not necessarily required 
as for massive halos. 

\noindent
2. The SF timescale $\tSF$ is defined as M$_g/\Psi$ with $\Psi$ 
as the SFR that, on the other hand,
in the self-regulated SF mode depends on $\rg^2$ \cite{koe95}.
Let me already emphasize here, that the theoretical treatment 
and modelling of SF self-regulation has to allow for the stellar 
energy of radiation and winds by massive stars already released 
in SF regions during their lives, i.e. prior to the explosion 
of supernovae typeII (SNeII). 
This necessity becomes clear when one continues a high SFR 
conditioned by gas infall and unaffected for a few million years 
until the first SNeII emerge. Since lower galaxy masses lead 
to less dense gas, SF is stretched over time for DGs. 

\noindent
And 3. 
As SF couples to stellar energy release, and since the counteracting
cooling process depends on $\rg^2$, the gas expands due to 
pressure support and reduces the SFR so that the effect 
of SF self-regulates non-linearly. 

Another important effect that seems to affect the whole network
of galaxy formation and evolution is ionizing radiation from the
first cosmological objects (supermassive stars, black holes, 
galaxies). 
Due to the re-ionization of the gas in the universe, its 
thermodynamical state is changed so that its accretion onto 
low-mass objects was reduced \cite{dij04} and gas already 
caught in minihalos evaporated again \cite{bar99}. 
Since massive objects remained almost unaffected by the 
re-ionization phase, while DGs should have
experienced delayed SF \cite{noe07}, this evolutionary dichotomy 
is observed as downsizing \cite{cow96}. 
Nevertheless, the assumption that all DGs were affected in the 
re-ionization era and in the same way would request overlapping 
Stroemgren bubbles in an almost uniformly ionized Universe. 
This, however, must be questioned and is contrasted by the existence 
and amplification of cosmological density structures \cite{par10}.

Another possible paths of DG origin is the formation of SF 
density enhancements in the tidal tails of merger galaxies 
\cite{duc07}. Since they should be free of DM and it is not yet 
well understood whether their SF acts in self-regulation 
the survival probability of these galaxy-like entities needs to
be explored \cite{rec07b}.

\section{
Dwarf irregular galaxies and their extreme evolutionary stages}
\label{dIrrs}

dIrrs are characterized by large gas fractions, ongoing SF, 
and low metallicities. That dIrrs contain the same 
or slightly higher gas fractions than giant spiral galaxies 
and mostly suffer the same SF efficiency, but appear with 
lower metallicity $Z$ than spirals, cannot be explained by 
simple evolutionary models. 
When gas is consumed by astration but replenished partly by 
metal-enhanced stellar mass loss, the general analytical 
derivation relates the element enrichment $Z(t)$ with the 
logarithm of decreasing remaining gas fraction 
$\mu = M_{\rm g}(t)/M_{\rm g}(0)$ as
$Z(t) = y\, [-ln(\mu)]$, where $y$ as the slope is 
determined by the stellar yield 
(see e.g.\ textbooks like \cite{pag10} or reviews as e.g. 
by \cite{pra08a,hen10}). 
As demonstrated by \cite{gar02} and \cite{zee01}, however, 
the effective yields of gas-rich galaxies decrease to 
smaller galaxy masses. This means that their element abundances,
particularly O measured in \HII regions, are smaller 
than those released by a stellar population and confined 
in a "closed box". 

Two processes can reduce the metal abundances 
in the presence of old stellar populations: 
loss of metal-enriched gas by galactic outflows or infall 
of metal-poor (even pristine) intergalactic gas (IGM).
It is widely believed, that a fundamental role in the 
chemical evolution of dIrrs is played by galactic winds,
because freshly produced metals in energetic events are
carried out from a shallow potential well of DGs through a 
wind (which will be therefore metal-enhanced).
Some SBDGs are in fact characterized by galactic winds 
\cite{mar95} or by large expanding supernova type II 
(SNeII)-driven X-ray plumes (e.g. \cite{hen98,mar02}). 
Studies have raised doubts to whether the expanding 
$\Ha$ loops, arcs, and shells mostly engulfing X-ray plumes, 
really imply gas expulsion from the galaxies 
because their velocities are mostly close to escape, 
but adiabatic expansion against external gas tends to 
hamper this. 

As an extreme, \cite{bab92} speculated that galactic winds 
are able to empty DGs from its fuel for subsequent SF  
and, by this, transform a gas-rich dIrr to a fading gas-poor 
system. In order to manifest this scenario and to study mass 
and abundance losses through galactic winds numerous 
numerical models are performed under various, but mostly 
uncertain conditions and with several simplifications 
(e.g. \cite{mlf99,str04}). 
The frequently cited set of models by MacLow \& Ferrara 
\cite{mlf99} (rotationally supported, isothermal \HI disks 
of dIrrs with fixed structural relations for four different gas 
masses between M$_{\rm g} = 10^6 - 10^9 \Msun$ and three 
different SNII luminosities in the center corresponding to 
SN rates of one per $3\times 10^4$ yrs to 3 Myrs) is mostly 
misinterpreted:  The hot gas is extremely collimated from the
center along the polar axis, but cannot sweep-up sufficient 
surrounding ISM to produce significant galactic mass loss.
On the other hand, the loss of freshly released elements from
massive stars is extremely high. Moreover, these models
lack of realistic physical conditions, as e.g. the existence of
an external pressure, self-consistent SFRs, a multi-phase
inhomogeneous ISM, and further more.

\begin{figure}
\vspace{-1.0cm}
\begin{tabular}{lr}
\resizebox{0.49\columnwidth}{!}{%
  \includegraphics{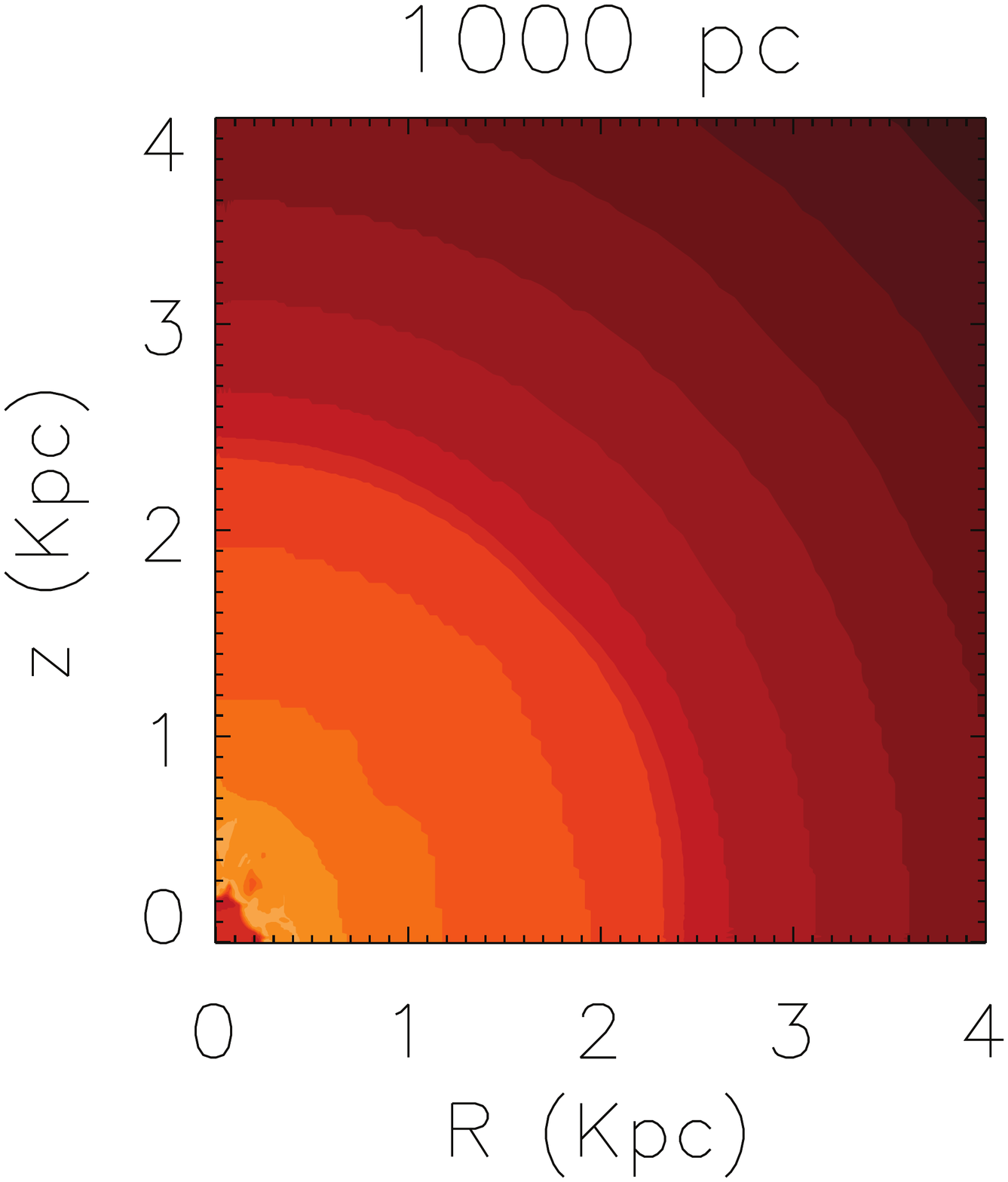} }
&  
\resizebox{0.49\columnwidth}{!}{%
\vspace{0.2cm}
\includegraphics{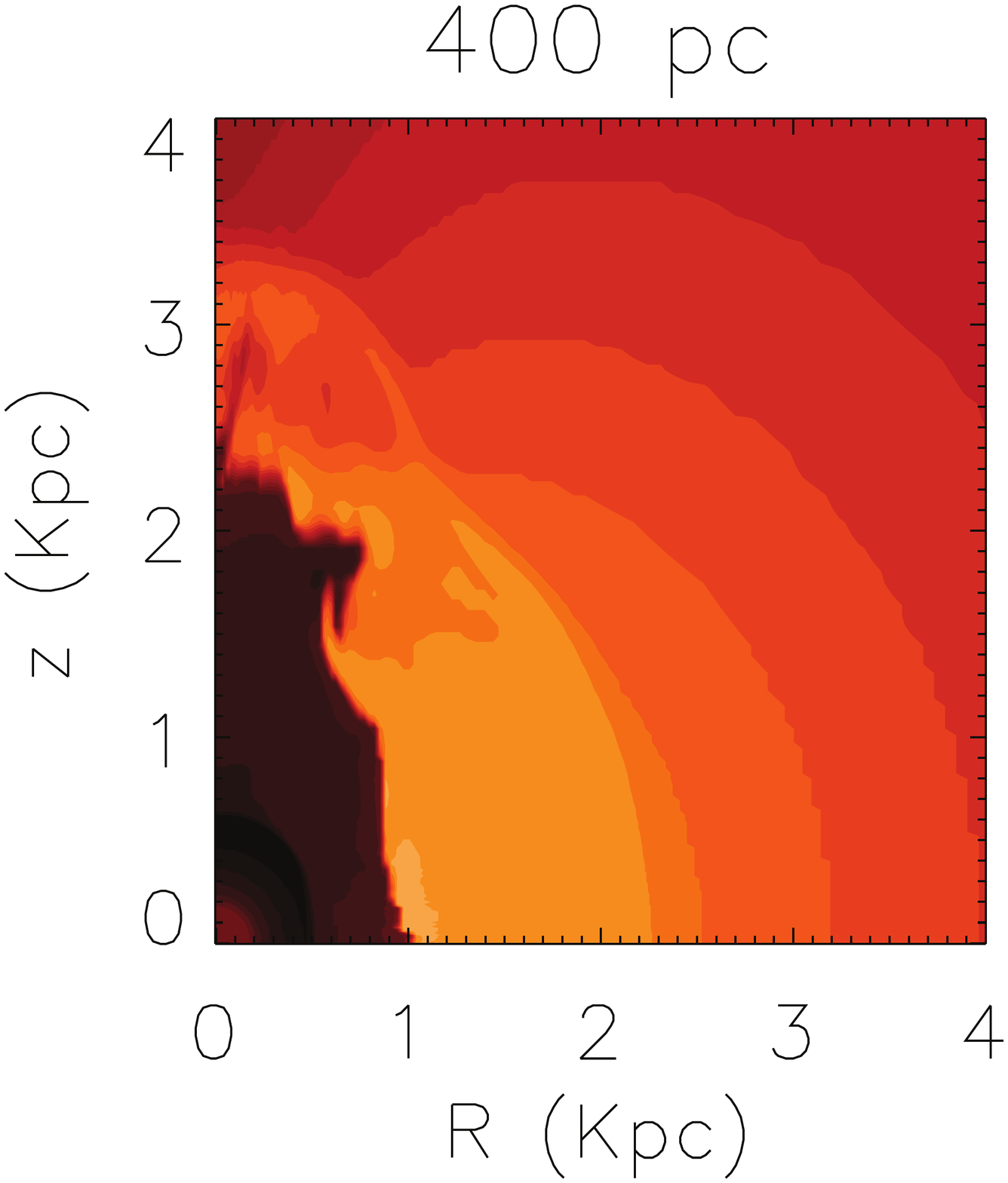} }
\end{tabular}
\vspace{-0.5cm}
\caption{
 Density contours after 200 Myr of evolution for models differing 
on the semi-minor axis b of their initial configurations 
(semi-minor axis indicated on top of each panel), while the 
semi-major axis is set to a=1 kpc:  
{\it left:} b=a=1 kpc (spherical); 
{\it right:} b=400 pc, (eccentricity of (a-b)/a=0.6). 
The density scale ranges from 10$^{-27}$ (black) 
to 10$^{-24} g/cm^3$ (brightest). 
{\it (for details see \cite{rec09}.)} 
}
\label{fig:wind}
\end{figure}

Also more detailed numerical simulations \cite{db99,rec06a}, 
show that galactic winds are not very effective in removing gas 
from a galaxy. Although galactic winds develop vertically, 
while the horizontal transport along the disk is very limited,
their efficiency depends very sensitively on the galaxy structure 
and ISM properties, as e.g. on the \HI disk shape \cite{rec09}.
Fig. \ref{fig:wind} reveals clearly that the more eccentric the disk 
is, the more pronounced does the superbubble expand.
On the one hand, the hot SN gas has to 
act against the galactic ISM, exciting turbulence and mixing
between the metal-rich hot gas with the surrounding \HI. Not
taken into account in present-day models is the porosity
of the ISM, consisting of clouds and diffuse less dense gas.
In particular, the presence of clouds can hamper the development 
of galactic winds through their evaporation. 
This so-called mass loading reduces the wind momentum and internal 
energy. Since the metallicity in those clouds are presumably lower
than the hot SNII gas, also the abundances in the outflow are 
diminished as e.g. observed in the galactic X-ray outflow of
NGC~1569 \cite{mar02} for which a mass-loading factor of 10
is derived to reduce the metallicity to 1-2 times solar.
In recent simulations \cite{sca10} demonstrate that turbulent 
mixing can effectively drive a galactic wind. Although they
stated that their models lead to a complex, chaotic distribution 
of bubbles, loops and filaments as observed in NGC~1569, 
other observational facts have not been compared.

Detailed numerical simulations of the chemical evolution of 
these SBDG by \cite{rec06b} could simultaneously reproduce 
both, the oxygen abundance in the warm gas as well as the 
metallicity in the hot outflow. Surprisingly, \cite{rec07a} 
demonstrated that the leakage of metals from a SBDG is not 
prevented by the presence of clouds because the clouds
pierce holes into the wind shells. This leads to a final 
metallicity a few tenths of dex lower than in models without 
clouds.
 
Consequently, the crucial question must be answered which 
physical processes trigger such enormous SFRs as observed 
in SBDGs and consume all the gas content within much less 
than the Hubble time. One possibility which has been favoured
until almost two decades ago was that at least some of these
objects are forming stars nowadays for the very first time.
Today it is evident that the most all metal-poor ones 
(like I~Zw~18) contain stars at least 1 Gyr old \cite{mom05},
and most SBDGs have several Gyrs old stellar populations.
This means that SF in the past should have proceeded in dIrrs, 
albeit at a low intensity and long lasting, what can 
at best explain their chemical characteristics, 
like for instance the low [$\alpha$/Fe] ratio \cite{lm04}.  
The [$\alpha$/Fe] vs. [Fe/H] behaviour is representative of
the different production phases, $\alpha$-elements from 
the short-living massive stars and 2/3 of iron from type Ia SNe 
of longer-living binary systems. If the SF duration in a galaxy 
is very short, type Ia SNe do not have sufficient time to 
enhance the ISM with Fe and most of the stars will be 
overabundant in [$\alpha$/Fe].

\begin{figure}
\vspace{-0.5cm}
\begin{tabular}{lr}
\resizebox{0.6\columnwidth}{!}{%
  \includegraphics[width=7cm]{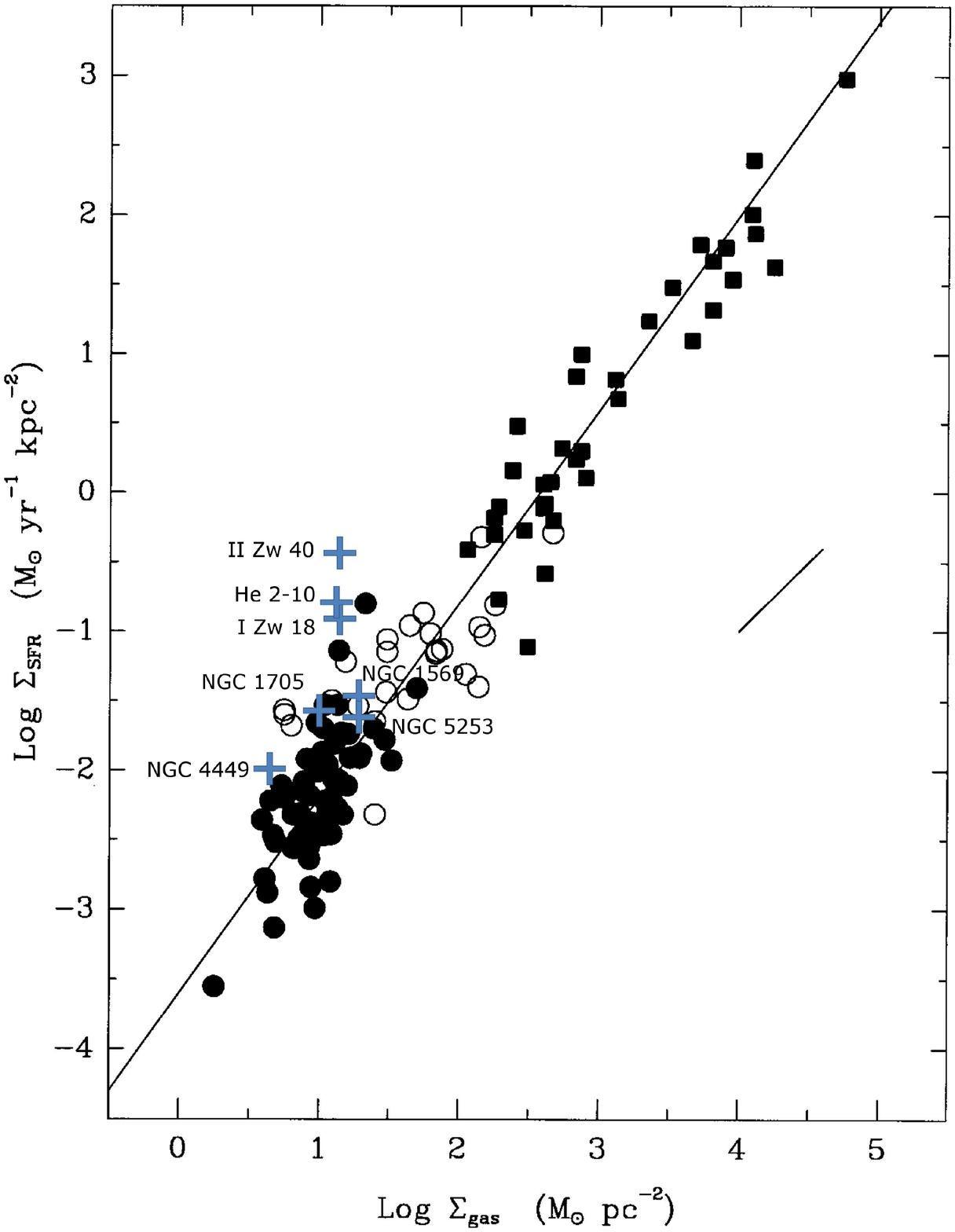} }
&
\begin{minipage}[t]{4.5cm}
\vspace{-9.5cm}
\caption{
 Comparison of the surface star-formation rate vs. \HI column
 density of a few prototypical starburst dwarf galaxies (SBDGs) 
denoted by crosses and names with the well-known Kennicutt-Schmidt 
relation derived by \cite{ken98} with an exponent of 1.4 
(long full-drawn line). 
The SBDGs' \HI surface density is averaged over the optical 
galactic body {\it (from \cite{kue11})}. 
} 
\end{minipage}
\hfill
\end{tabular}
\vspace{-0.5cm}
\label{fig:KS} 
\end{figure}

In most SBDGs large \HI reservoirs enveloping the 
luminous galactic body have been detected 
(NGC~1569 \cite{sti02},
NGC~1705 \cite{meu98}, 
NGC~4449 \cite{hun98}, 
NGC~5253 \cite{kob08}, 
I~Zw~18 \cite{zee98c}, 
II~Zw~40 \cite{zee98b})
with clearly disturbed gas kinematics and disjunct from the 
galactic body. Nevertheless, in not more than
two objects, NGC~1569 \cite{mue05} and NGC~5253 \cite{kob08}
gas infall is proven, while for the other cases the gas kinematics
obtrudes that the gas reservoir feeds the engulfed DGs. 
In another object, He~2-10, the direct collision with an intergalactic
gas cloud \cite{kob95} is obviously triggering a huge SB.
Reasonably, for their measurable \HI surface density the SFRs of
most of these objects exceed those of ''normal'' gas disks 
(Fig.\ref{fig:KS}).

Yet it is not clear, what happens to dIrrs if they move into a
region with increasing external pressure as e.g. by means of a
denser IGM and of ram pressure when they fall into galaxy clusters. 
In sect.\ref{dEs} we will discuss the
effect of ram pressure on the structure of the ISM for which
numerical models  for spiral galaxies (e.g. \cite{roe05})
as well as for gas-rich DGs (e.g. \cite{mor00}) exist, but only 
hints from observations (as e.g. for the Magellanic stream).
The effect on the SFR due to compression of the ISM is observed, 
but not yet fully understood and convincingly studied by models. 
\cite{cor06} e.g. observed a coherent enhancement of SF in 
group galaxies falling into a cluster.

\begin{figure}[h]
\hspace{0.5cm}
  \includegraphics[width=10cm]{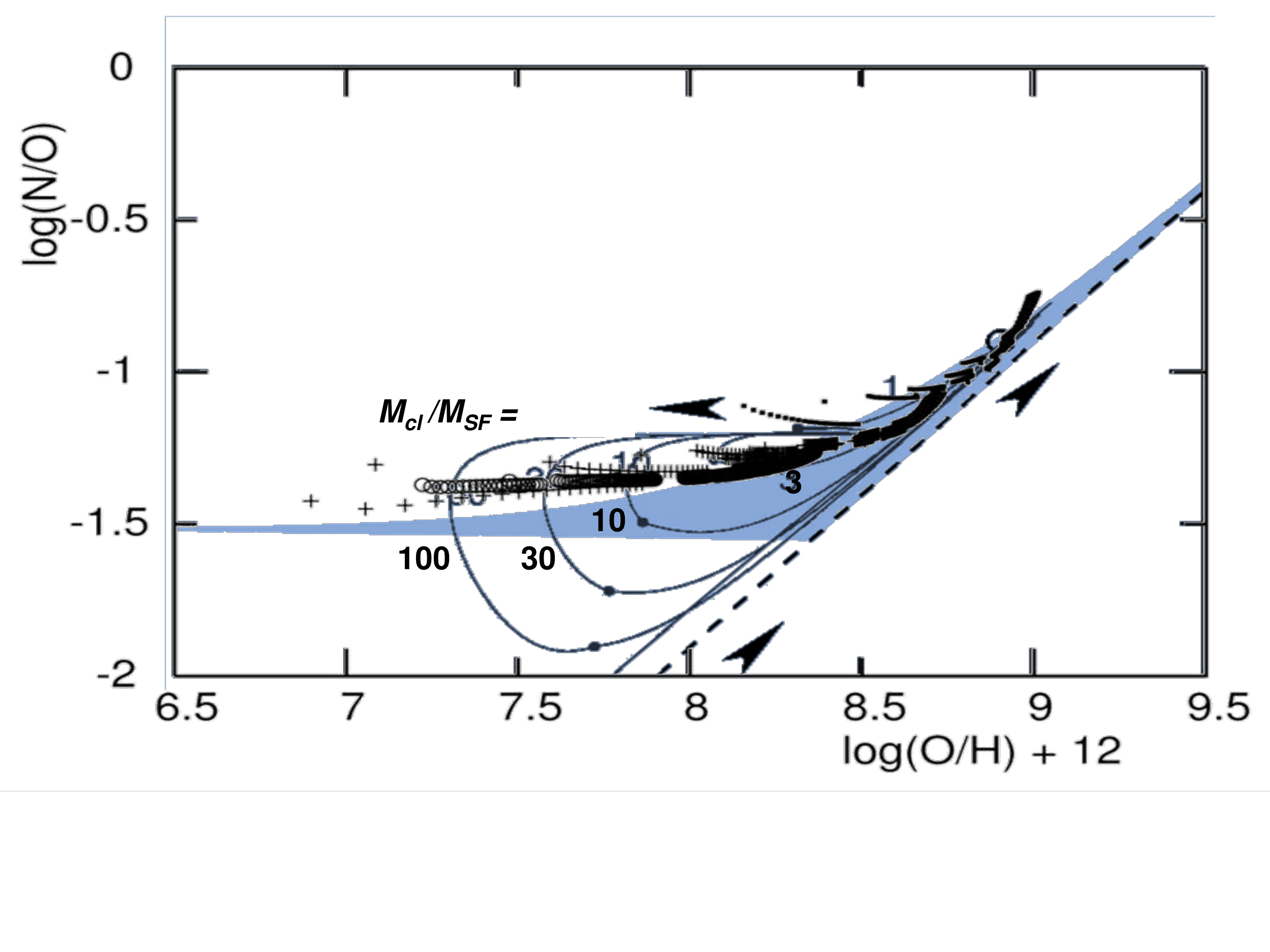} 
\vspace{-1.cm}
\caption{
The abundance ratio N/O as a function of oxygen abundance
observed in spiral and irregular galaxies (shaded area,
after \cite{zee98a}) overlayed with evolutionary loops due to 
infall of primordial intergalactic gas clouds. 
These have different mass fractions $M_{cl}/M_{SF}$ with 
respect to the mass involved in the SF region The crosses
represent evolutionary timesteps of models, the arrows depict
the direction of the evolutionary paths.
The dashed straight line represents a simple model relation  
for purely secondary nitrogen production. 
{\it For discussion see \cite{koe05}.} }
\label{fig:NO} 
\end{figure}

Although the mass-metallicity relation also holds for dIrrs
and even steepens its slope \cite{tre04}, what can be interpreted 
by galactic mass loss and the corresponding lower effective
yield \cite{gar02}, the abundance ratios are unusual. As
mentioned above, O/Fe reaches already solar values for subsolar
oxygen or iron abundances. While this can be explained
by a long SF timescale, another characteristic signature 
is that the ratio \lNO stays at about -1.6 to -1.5 
with O abundances below 1/10 solar and with an increasing
scatter with O (see Fig.\ref{fig:NO}). 
Their regime of N/O--O/H values overlaps with those of 
\HII regions in the outermost disk parts of spirals at about 
\OH = 8.0 ... 8.5 \cite{zee98a}. 

In the 90th several authors
have tried to model these observations by SF variations with
gas loss through galactic winds under the assumptions that these
dIrrs and blue compact DGs (BCDs) are young and experience their 
first epochs of SF (for a detailed review see \cite{hens99}). 
Stellar population 
studies contradict this youth hypothesis, so that another 
process must be invoked. Since these objects are embedded into 
\HI envelops and are suggested to suffer gas infall as manifested
e.g. in NGC~1569 (see above, \cite{sti02,mue05}), the influence of 
metal-poor gas infall into an old galaxy with continuous SF on 
particular abundance patterns were exploited by Koeppen \& Hensler
\cite{koe05}. 
With the reasonable assumption that the fraction of 
infalling-to-existing gas mass increases with decreasing galaxy 
mass, their results could match not only the observational regime of 
BCDs in the [\OH]-\lNO space but also explain the shark-fin shape 
of observational data distribution \cite{koe05}.

Fig.\ref{fig:NO} demonstrates how self-enriched galaxies which
have reached the secondary nitrogen-track already within 
2-3 Gyrs of their evolution are thrown back in O abundance
by gas infall while N/O stays the same. After a time delay depending
on the mass fraction of infalling gas to that existing already 
within the SF site, along a loop-like evolutionary paths in the 
[\OH]-\lNO diagram the ISM abundances reach again the starting
point. In summary, one can state that old dIrrs {\it mutate} 
temporarily to youngly appearing examples with respect to their 
gas abundances.

\section{Dwarf elliptical galaxies}
\label{dEs}

dEs are frequently denoted as examples of ``stellar fossile'' 
systems in which the bulk of their SF occurred in the past.  
They are preferentially located in morphologically evolved 
environments \cite{tre02}, i.e. in regions with high 
galaxy densities and dominate the morphological types of galaxies 
in clusters, as e.g. Virgo, Coma, Fornax, and Perseus. 
Furthermore, dEs cluster strongly around luminous elliptical/S0 
galaxies \cite{tul08}. The evolution of this galaxy type
should be mainly caused by gas and tidal effects on SF and 
structure and indicates that it is strongly affected by environment. 

Already \cite{bot85} found that cluster dEs are usually almost 
free of interstellar gas and contain few young stars.  
In trying to understand the dE population, structural regularities 
and correlations must be studied, as it is known since the 80th, 
between optical surface brightness and luminosity 
\cite{kor85,fer94} 
and between luminosity and stellar velocity dispersion  
which also correlates with metallicity (e.g. \cite{pet93}).
Boselli et al. \cite{bos08b} proposed to understand these 
''Kormendy'' relations by processes having transformed dIrrs after 
their cluster infall, but accuse the still existing lack of 
numerical simulations of such transformation.
  
Furthermore, dEs often have flattened profiles but are mostly
kinematically supported by their stellar velocity dispersion 
rather than by rotation. 

The combination of low gas-mass fractions and moderate-to-low 
stellar metallicities in dE (about 0.1 of solar or less) is 
a key feature of this class. 
Their lower stellar abundances \cite{haa97} suggest that 
extensive gas loss occurred during their evolution and SF ceased 
due to a lack of raw material rather than exhaustion of the gas 
supply through SF. 
Galactic winds are therefore a hallmark of modern models for dE 
evolution, starting from the basic consideration by \cite{lar74} 
and continued with the study by \cite{ds86}. They are commonly 
assumed to have cleaned out DGs soon after their formation. 
As mentioned in sect. \ref{dIrrs}, however, gas expulsion by means 
of galactic winds is inefficient from our understanding of
the multi-phase ISM
and requires even in low-mass systems a DM-to-baryonic matter 
ratio \cite{mlf99} much smaller than assigned to DGs in the 
classical formation picture (e.g. \cite{mat98}). 

There are two competing scenarios for the origin of cluster dEs. 
On the one hand, those low-mass galaxies are 
believed to constitute the building blocks in $\Lambda$CDM 
cosmology and should therefore have evolved congruently with 
the mass accumulation to more massive entities, galaxies
and galaxy clusters. For those, orbiting in a cluster the stellar
component must be heated continuously by harrasment of more 
massive cluster galaxies and thus be pressure supported.
On the other hand, a variety of observations are available 
which also support discrepant scenarios of dEs evolution.   
Recent \HI studies of Virgo cluster dEs \cite{con03} 
and also those of the Fornax cluster (see e.g.\ \cite{mic04})
have unveiled that a small but significant fraction of them 
contains gas, has experienced recent SF, and can be argued 
from internal kinematics and cluster distribution data to represent 
an infalling class of different types of gas-rich galaxies in 
or after the state of morphological transformation. Further findings 
of a significant fraction of rotationally supported dEs in the 
Virgo cluster \cite{zee04} and also disk features as e.g.
spiral arms and bars \cite{lis06} support the possibility 
of morphological transformation from gas-rich progenitor DGs to 
dEs thru gas exhaustion. Boselli et al. \cite{bos08a} have 
comprehensively discussed the different processes of dE origin.

A separation into dE subclasses with respect to their origin
should also be visible in an intermediate-age stellar population,
blue centers, and flatter figure shape. 
Indeed,  dEs in the Virgo cluster can be divided into different 
subclasses \cite{lis07} which differ significantly 
in their morphology and clustering properties, however, do not show 
any central clustering, but are distributed more like the late-type 
galaxies. These types of dEs show different disk signatures,
such as bars and spiral structures, are not
spheroidal, but rather thick disk-like galaxies.  
Similar shapes were also found for the brighter, non-nucleated dEs.
There is only a small fraction of nucleated dEs whithout any 
disk features or cores, which keep the image of spheroidal objects 
consisting of old stars.

\begin{figure}[h]
\vspace{-1.0cm}
\begin{tabular}{lr}
\resizebox{0.6\columnwidth}{!}{%
  \includegraphics[width=7.5cm]{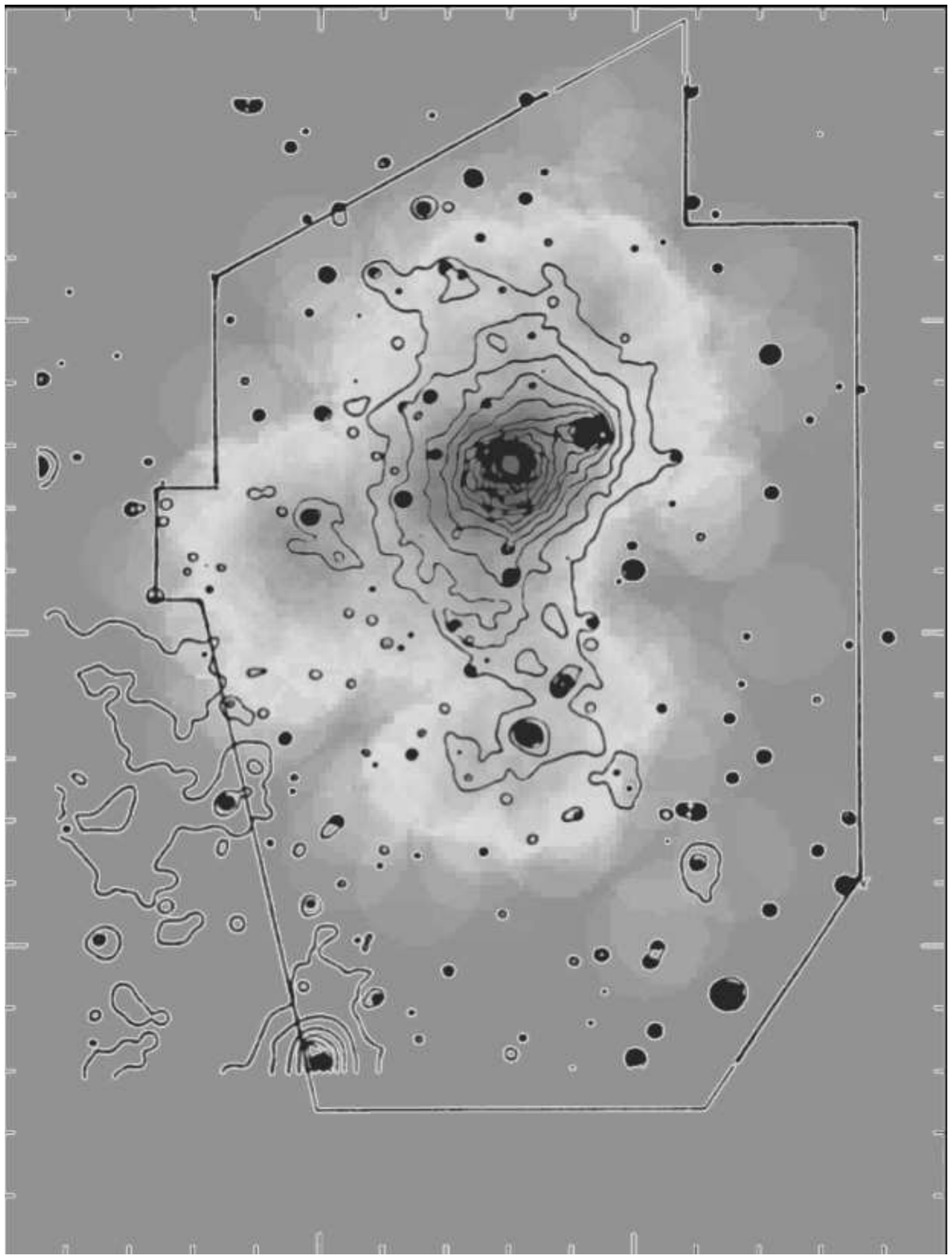} }
&
\begin{minipage}[t]{4.5cm}
\vspace{-8.7cm}
\caption{
Distribution of dEs in the Virgo Cluster divided into 
rapid (white) and slow objects (darker central part) 
overlaid with X-Ray brightness contours 
{\it (courtesy by Thorsten Lisker)}.
}
\end{minipage}
\hfill
\end{tabular}
\vspace{-1.0cm}
\label{fig:VCdE} 
\end{figure}

A figure analysis of Virgo dEs correlates with the averaged 
orbit velocity in the sense that flatter  
dEs show on average a larger orbital velocity (700 km/s) than 
those originating within the cluster (300 km/s) \cite{lis09} 
(see Fig. 4). 
This kinematical dichotomy is expected 
because galaxies formed in virial equilibrium within the cluster
retain their initial kinetic energy while the cluster mass grew. 
Galaxies falling into the present cluster potential must 
therefore possess larger velocities.
To obtain information about both evolutionary stages, 
the young infalling vs. the late cluster members, \cite{got09} 
studied SDSS data. The basic model is that dIrrs which are formed 
outside the Virgo Cluster and becoming stripped on their infall,
by this being transformed into dEs, should reveal properties 
recognizably different from dEs which have already aged in the 
cluster, as e.g. colors, effecitve radius, radial stellar 
distribution, and abundances. One result by \cite{got09} is that 
for the two dE populations, with and without cores, distinguished 
by their Sersic parameter, there is only a slight indication 
that non-nucleated dEs are more concentrated towards the inner 
cluster regions, whereas the fraction of nuclated dE is
randomly distributed, while \cite{lis07} found it to increase
with distance. 
An analysis of the relation between the central surface brightness
and the Sersic parameter shows the expected tendency to higher 
values for brighter galaxies. Furthermore, there were no further
relations found of the Sersic parameter, the effective radius, 
or the distance from M~87.

Deeper insights are provided by spectra. 
Koleva et al. \cite{kol09} found 
most dEs in the Fornax cluster to be roundish and to contain 
significant metallicity gradients already in the old stellar
population. They argue that this is due to a lack of sufficient 
mixing. In contrast, rotationally supported dEs have flat 
metallicity distributions. Compared with simulations this 
can be attributed to galactic winds, but the picture of 
metallicity and gradients is not yet clear. 
While \cite{spo09} show a tight positive correlation between 
the total metallicity [Z/H] and the mass, \cite{kol09}
do not find any trend involving [Fe/H] for Fornax-cluster and 
nearby-group dEs. 

Moreover, from the deconvolution of the SF history of their 
sample dEs with respect to the central 1 arcmin and within 
the effective radius \cite{kol09} draw the conclusions that 
for a few objects SF episodes occurred in the very center 
even within the last 1 Gyr.
From a systematic study of the central Fornax-cluster dEs' dynamics 
\cite{rij10} conclude that these objects stem from an infall 
population of late-type DGs and has been transformed to dEs 
by ram-pressure stripping (RPS). 

Toloba et al. \cite{tolo09} derive for Coma cluster dEs to be 
weaker in carbon than dEs in low-density environments, 
while similar in nitrogen. Actually, they \cite{tol11} 
also find that pressure supported Virgo dEs show higher dynamical 
mass-to-light ratios than rotationally supported dEs of 
similar luminosity and further that dEs in the outer parts 
of the cluster are mostly rotationally supported with disky 
shapes. 
Rotationally supported dEs even follow the Tully-Fisher relation. 
One fundamental and most spectacular result \cite{tol11} is, 
however, that dEs are not DM-dominated galaxies, at least 
up to the half-light radius.  

Correlations of both signatures, SF history and metallicity
gradients, for cluster-member dEs vs. infall dEs should be
derived for more clusters, but observations are unfortunately 
very time-expensive if possible at all.

In addition to classical dEs, ultra-compact DGs (UCDs) have 
been detected and classified as a new type of cluster dEs 
that differ by their intrinsic structure and brightness 
(see e.g. \cite{phi01,gre09}). 
The origin of these peculiar DGs is mysterious and not yet 
understood but requests transformation if they are surviving 
nuclei of tidally stripped nucleated DGs \cite{goe08}.

\section{Morphological transitions}

As the transformation picture from late-type dwarfs into dEs is 
still not completely understood and only qualitative, 
in sect.\ref{dEs} we tried to shed light on the expected 
witnesses of differences in system parameters which allow to 
distinguish between two different populations of dEs. 
That almost all DGs can be associated with morphological types and 
that only a few exceptions show morphological transformations, 
implies that the act of mutation seems to happen rapidly and 
thus to be observable with only low probability. 

During the approach to galaxy clusters, ram pressure should act on
dIrrs already at group conditions or in the outskirts of clusters 
\cite{mar03, roe05}. When this process pushes the gas out of dIrrs, 
wouldn't one expect to observe many head-tail structures of stellar 
body vs. stripped-off ISM? Several candidates exist which, however, 
are characterized as BCDs with decentered bright star-forming knots 
(see e.g. \cite{noe00}). 
The recently detected best candidate in the rapid phase of RPS 
in the Virgo cluster is VCC1217/IC4318 \cite{fum11}: 
A main almost old stellar body of $3 \times 10^8 \Msun$
leaves behind a bundle of $\Ha$ and near-UV emitting knots.
Nevertheless, the gas distribution and SF progression within some 
blobs are not fully understandable. 
It is still, in general, unclear to what extent the ram pressure 
can trigger SF by compression of the ISM. 

Other BCDs are observed although their gas is already exhausted 
\cite{del08}. Peebles et al. \cite{pee08} find a number of 
dIrrs with excessively high 
metallicity, what they interpret as the last stage of gas 
consumption before they reach the dE state.

On the other hand, rejuvenation of a fraction of cluster dEs 
seems also to occur which are found to harbour central warm gas 
\cite{mic04}. Whether this fact is indicative of a possible
re-transformation from gas-free to gaseous DGs by gas 
accretion is a matter of debate, since it seems impossible
within cluster environments, but plausible in less dense regions 
where gas infall enhances SF or even triggers starbursts 
(see sect.\ref{dIrrs}) and should not only affect dIrrs.

\section{Dwarf spheroidal galaxies}
\label{dSphs}

The possible cycle of morphological mutations, i.e. from gas-rich
objects to gas-poor systems by means of gas expulsion and back to
a significant gas content by gas accretion, can be explored
in the local environment, namely, in the galaxies around 
the Milky Way, their satellites. Except the Magellanic
Clouds most of these can be characterized as gas-free spheroidal 
systems which manifest the faint end of dEs. 
Since these dSph are gathered around 
massive galaxies like our MWG and M31 orbiting them as satellites,
go down to the lightest and most metal-poor end of galaxies, 
they have attracted increasing attention over the last years 
with the advent of more advanced observing facilities. 
Understanding their formation and evolution is of substantial 
relevance for our astrophysical picture of cosmological structure
formation and of galaxy evolution. 
Four main questions are addressed: 
\vspace{-0.2cm}
\begin{enumerate}
\item How and when did they form? They all harbour a very old stellar 
population \cite{tol09} and, therefore, seem to have been unaffected 
by the re-ionzation era \cite{gre04}.
\item Is their existence as satellite system typical for all massive galaxies? Their origin and DM content is still questioned 
by some authors \cite{kro10} because of the large discrepancy of
the number of objects really observed vs. expected from
$\Lambda$CDM cosmology and because of their orbit concentration 
to the so-called disk-of-satellites, also found around M~31. This 
invokes the preference of their tidal-tail origin \cite{met09}.
The observed large velocity dispersions, which are otherwise 
applied as representative to derive the M/L ratio is then caused 
by tidal effects. 
\item How is their evolution determined by the vicinity of the 
massive mature galaxy? Not only the tidal field must have a 
disruptive effect, but also a gaseous halo of the central galaxy 
will interact with the ISM of the dSphs \cite{may07}.
That the relation of the gas fraction bound to the dSphs is 
increasing with distance from the MWG \cite{har01}, points into 
that direction.
\item Vice versa the question arises, how do the satellites influence 
the structure and evolution of the mature galaxy, here the MWG.
\end{enumerate}
  \vspace{-0.2cm}

\noindent
The first three questions also concern the morphological transition
from gas-rich satellites to dSphs. Nevertheless, dSphs follow 
a mass-metallicity relation \cite{dek03,gre03}
and continue the total brightness vs. central surface-brightness 
relation from normal dEs to the faint end \cite{gre03}.

As the first models, Hensler et al. \cite{hen04} performed 
\cd\ simulations \cite{hen03} of spherical low-mass galaxies 
in order to study galaxy survival, SF epochs and rates, gas loss, 
and (final) metallicity.
They demonstrate that due to the SF self-regulation only short 
but vehement initial SF epochs occur and lead to mass-dependent 
gas loss.
Nonetheless, the DGs remain gravitationally bound with the further
issue that more cool gas survives than it is observed, but it forms 
a halo around the visual body. Although the stellar energetic 
feedback is the driving mechanism to expel the gas, its effect is
not as dramatic as obtained in semi-analytic models \cite{sal08} 
and the amount of unbound mass is considerably lower. 
To get lost, this gas has to be stripped off additionally 
\cite{gre03} what probably happens by means of ram pressure 
of the galactic halo gas \cite{may07} or by tidal stripping 
\cite{rea06}. 
Otherwise it can return to the DG and produce subsequent events,
from a second SF epoch to SF oscillations. 
The external gas reservoirs around some dSphs \cite{bou06}, in
particular also the \HI that is kinematically coupled with
the Scl dSph, might witness this effect.

The fascinating wealth of data and their precision on stellar ages
and kinematics, on their chemical abundances, abundance gradients, 
and tidal tails of dSphs 
(for most recent reviews see e.g. \cite{koc09} and \cite{tol09}) 
have triggered numerous numerical models. 
Although  they are advanced since \cite{hen04} to 3D hydrodynamics
(see e.g. \cite{mar08} and \cite{rev09}), they still lack of a
self-consistent treatment of both, internal processes, as e.g. 
SF self-regulation (see sect.\ref{intro}), and the environmental
influences as e.g. tidal effects, external gas pressure, 
gas inflow, etc. 

In a recent paper \cite{rev09}, e.g. a large set of DG models 
is constructed with the method of 
smooth-particle hydronamics (SPH), but considers all of them 
in isolation. In most of their models sufficient gas mass 
is retained and can fuel further SF epochs, if it would not be
stripped of by ram pressure or tidal forces, as the authors 
mention. Those models that fit the presently best studied 
dSphs Fnx, Car, Scl, and Sex, are than chosen as test cases 
for further exploration.
Although their results do not deviate too much from the further
observational data, in addition to the already mentioned 
neglections, three further caveats exist:
1) If models are selected according to any agreement with 
one or two observed structural parameters, it is not surprising 
if also other values would not deviate significantly.
2) The numerical mass resolution of the SPH particles is 
too low to allow quantitative issues of galactic winds, 
heating and cooling, etc.
3) Because of the single gas-phase description released metals 
are too rapidly mixed with the cool gas and the metal-enrichment 
happens too efficiently. Despite these facts, with appropriate 
initial conditions always models in agreement to observations 
can be found.

\begin{figure}
  \includegraphics[width=12cm]{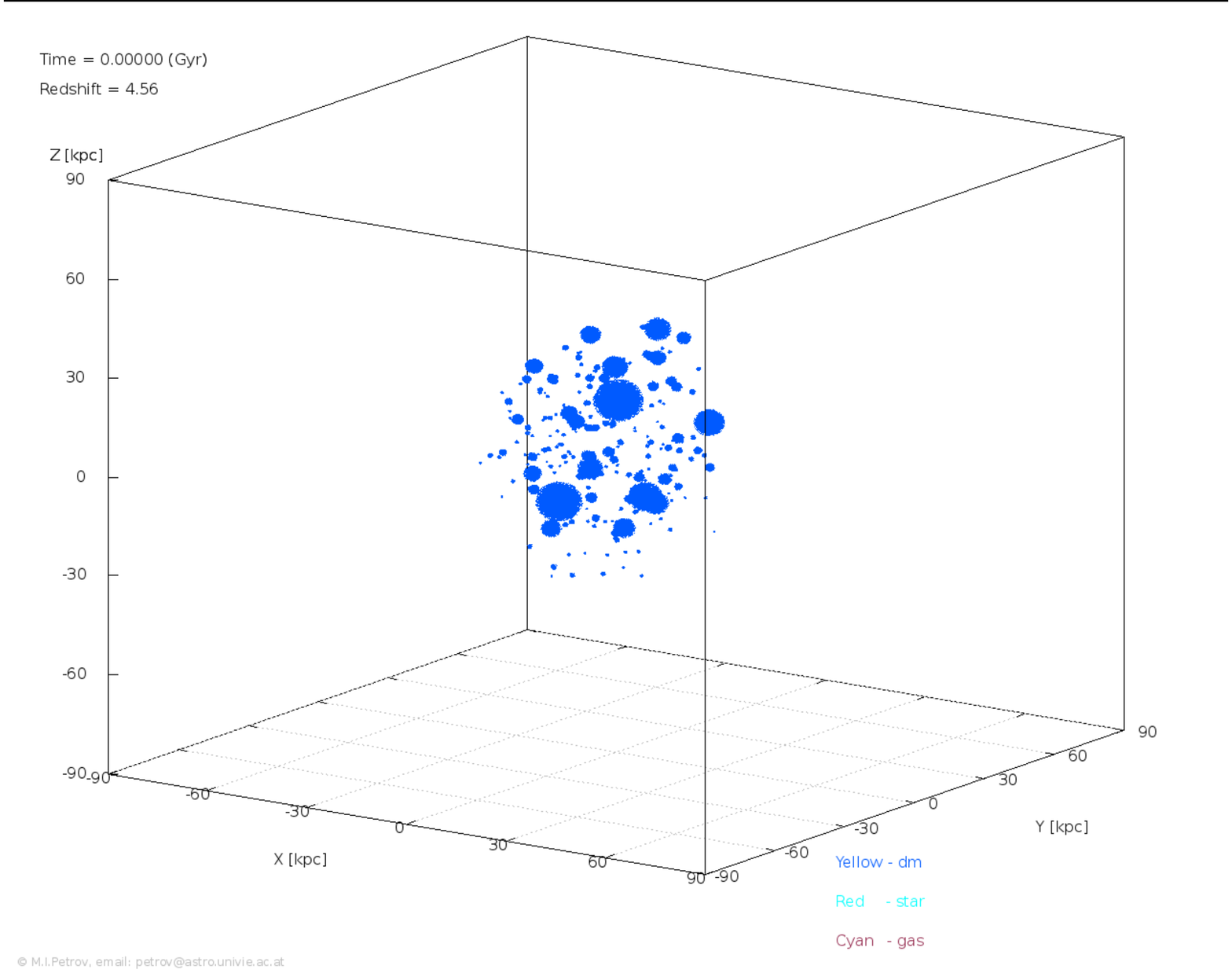}  
  \includegraphics[width=12cm]{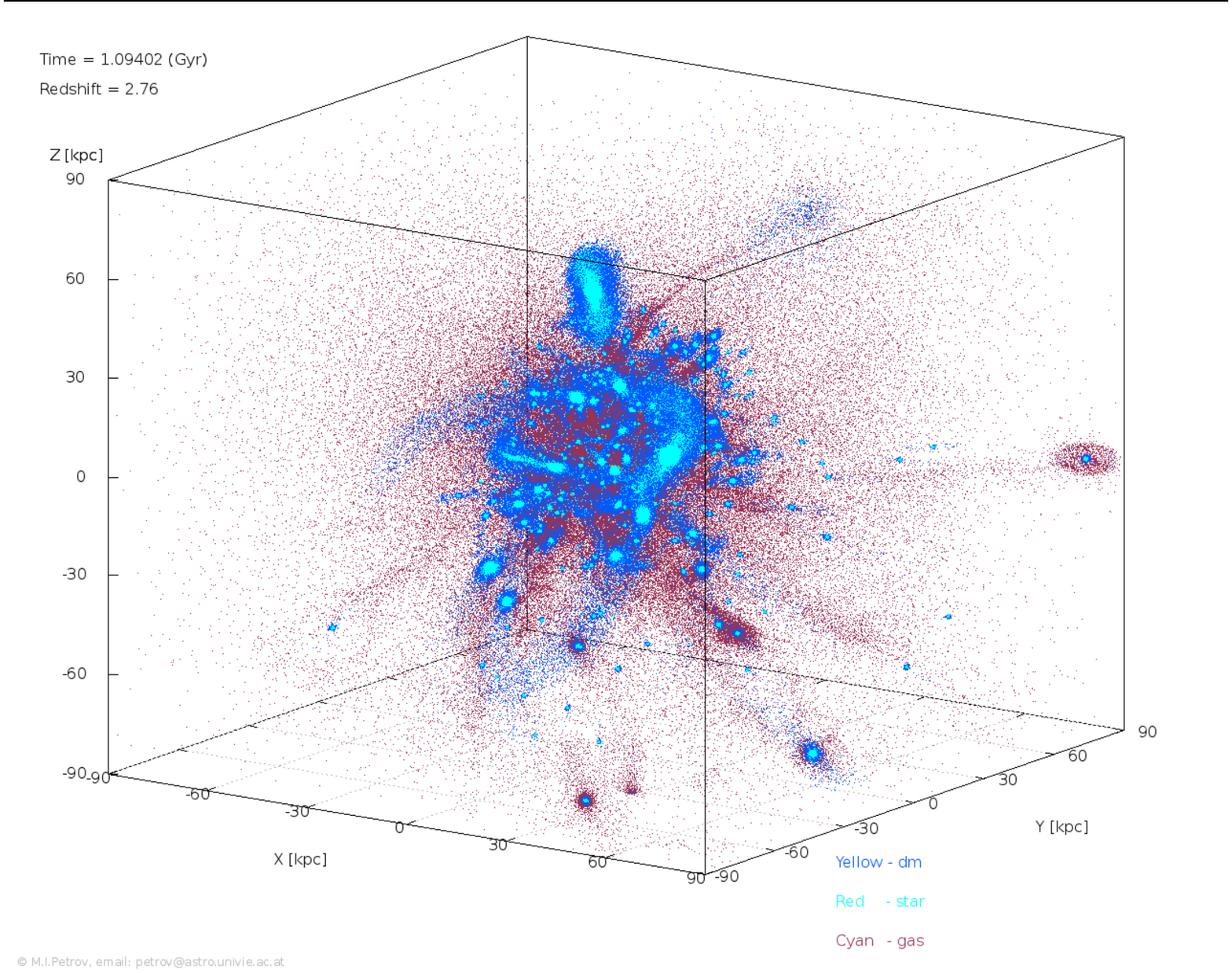} 
\caption{
Cubes of 200 kpc length around the Milky Way (at their center). 
{\it upper panel:} 
Initial conditions of the Milky Way's satellite system:    
Distribution of Dark Matter (DM) subhalos within a sphere 
of 40 kpc radius around the Milky Way at redshift z=4.56. 
{\it lower panel:} 
Snapshot of the satellites' dynamical evolution 
1 Gyr after the numerical onset, i.e. at redshift z=2.76. 
The DM subhalos are filled with baryonic gas of mass fraction 
of 17\%, form stars, and lose mass of all constituents due to 
tidal interactions among the satellite system. 
For discussion see text. 
}
\label{fig:sat}
\end{figure}

Although the advancement to a two-phase ISM treatment in SPH is 
not trivial and implies various numerical problems, but is 
not impossible \cite{ber03,har06,sca06}, such treatment 
would be absolutely necessary in order to approach reality and
to achieve reliable results.  In addition, the \cd\ interaction 
processes must be applied \cite{ber03}.

From the $\Lambda$CDM structure-formation paradigm and from
numerical simulations with different computational tools,
subhalos are expected to assemble around massive halos and to
accumulate their masses. If they have already experienced SF, 
their stars should be merged into a spheroid and be identifiable 
by their kinematics and chemical abundances. 
Although these low-mass subhalos, their baryonic content as
dSphs, and the accretion scenario \cite{joh08}, therefore, 
serve as the key to pinpoint this cosmological paradigm, 
observational detections of stellar streams within the 
Milky Way (MW) halo are rare \cite{sea08}. 
Furthermore, the stellar abundances 
in present-day dSphs deviate mostly from the halo, in particular 
$\alpha$/Fe with Fe/H, which characterizes the SF timescale 
(\cite{tol09} and e.g. for the Car dSph \cite{koc08}).

Yet evolutionary models of dSph with respect to SF, chemistry, 
and gas expulsion and their comparison with the Milky Way halo
are still too simplistic. 
While their accretion epoch occurred continuously over 
the Hubble time some models \cite{pra08b} only considered it as
a short early event; their gas is not only removed by tidal 
stripping and RPS \cite{may07} but also re-accreted on their
orbits around the MW \cite{bou06}; in general, consideration 
as isolated systems lacks reality \cite{rev09}.

To model instead of isolated subhalos the evolution of the system 
of dSphs in the gravitational field of the MWG for which the accretion 
by the host galaxy is probable over the Hubble time, the Via Lactea II
 \cite{die08} simulation was used. Since an acceptible computational 
time limits the  number of gas particles to two million as also 
for the DM and in order to reach a high mass resolution 
of 10$^3 \Msun$ per SPH particle, only 250 subhaloes as 
DM progenitors of dSphs in the mass range of 
$10^6 < M_{sat}/\Msun < 6\cdot10^8$ from $z=4.56$ could be followed.
Unfortunately, this fact limits the radius of consideration to 
within a radius of 40 kpc around the MW's center of mass.
In order to study the construction of the MWG halo by accretion
of subhalos including baryonic matter, both gas and stars, as a 
first step, the \cd evolution of the dSph system is followed for 
the first Gyr, i.e. until redshift $z=2.76$ \cite{ph11} 
(see Fig.\ref{fig:sat}).
For the simulations an advanced version of the single-gas 
chemo-dynamical SPH/N-body code  is applied, treating
the production and chemical evolution of 11 elements. 

Starting with a $10^4$ K warm gas of 17\% of the subhalo masses
in virial equilibrium and under the assumption that 
re-ionization is improbable to have affected the Local Group 
dSphs \cite{gre04}, cooling allows the gas particles to achieve 
SF conditions in all satellites, but its efficiency directly 
depends on the mass of a satellite and its dynamical history 
(merging with other satellites or disruption by the MW gravitational
potential). 
The stellar feedback by SNeII releases sufficient energy to expel 
hot gas from the main bodies of less massive dSphs, facilitated 
by tidal interactions. This gas accumulates in the MW halo 
while massive dSphs merge and continue SF. 
For the first $10^8$ yr of the simulation there is a considerable
variance of stellar oxygen abundance in the whole system 
$(-5. \leq [O/H] \leq -0.5)$ reflecting the very inhomogeneous 
production and distribution of enriched gas.
After $10^8$ yrs the merging of satellites' ISM promotes the mixing 
of heavy elements. Finally, almost completely recycling of the gas 
erases the abundance inhomogeneities so that O in stars converges 
to $-1. \leq [O/H] \leq 0.$ with a small dispersion.

Detailed analyses of the SF history, gas exchange, stellar abundance 
evolution of dSphs and the MW halo in the early universe are
presented in a comprehensive paper \cite{ph11} and will be discussed 
with their implications for our cosmological picture. 

\begin{acknowledgement}
The author is grateful to Alessandro Boselli, Joachim Koeppen, 
Thorsten Lisker, Polis Papaderos, Mykola Petrov,  Simone Recchi, 
and further more for their contributions and continuous discussions 
to this field and to the referee for valuable comments.
\end{acknowledgement}
\vspace{-0.5cm}

\input{hensler.referenc}

\end{document}

%% file: hensler.referenc.tex
%
%
%

%% file: hensler.jenam2010.bbl
\begin{thebibliography}{99.}
%
%
%
%

\bibitem{bab92}
   Babul, A. \& Rees, M.J. 1992, MNRAS, \textbf{255}, 346

\bibitem{bar99}
   Barkana, R. \& Loeb, A. 1999, ApJ, \textbf{523}, 54

\bibitem{ber03}
   Berczik, P., Hensler, G., Theis, C., \& Spurzem, R. 2003,
   Astrophys. Space Sci. \textbf{284}, 865

\bibitem{bos08a} 
   Boselli, A., Boissier, S., Cortese, L., \& Gavazzi, G. 
   2008a, ApJ, \textbf{674}, 742

\bibitem{bos08b} 
   Boselli, A., Boissier, S., Cortese, L., \& Gavazzi, G. 
   2008b, A\&A, \textbf{489}, 1015

\bibitem{bot85}
   Bothun, G.D., Mould, J., Wirth, A., \& Caldwell, N. 
   1985, AJ, \textbf{90}, 697

\bibitem{bou06}
   Bouchard, A., Carignan,S C., \& Staveley-Smith, L. 
   2006, AJ, \textbf{131}, 2913

\bibitem{con03}
  Conselice, C.J., O'Neil, K., Gallagher,J.S. \& Wyse, R.F.G. 
  2003 ApJ, \textbf{591}, 167

\bibitem{cor06}
   Cortese, L., et al. 2006, A\&A, \textbf{453}, 853

\bibitem{cow96}
   Cowie, L.L., Songaila, A., Hu, E.M., \& Cohen, J.G. 
   1996, AJ, \textbf{112}, 839

\bibitem{ds86} 
   Dekel, A., \& Silk, J. 1986, ApJ, \textbf{303}, 39

\bibitem{dek03}
   Dekel, A. \& Woo, J. 2003, MNRAS, \textbf{344}, 1131

\bibitem{dek09}
   Dekel, A., et al. 2009, Nature, \textbf{457}, 451
 
\bibitem{del08} 
   Dellenbusch, K.E., Gallagher, J.S., Knezek, P.M., \& Noble, A.G.
   2008, AJ, \textbf{135}, 337

\bibitem{db99}
   D'Ercole, A., \& Brighenti, F. 1999, MNRAS, \textbf{309}, 941

\bibitem{rij10}
   De Rijcke, S., Van Hese, E., \& Buyle, P. 
   2010, ApJ, \textbf{724}, L171

\bibitem{die08}
   Diemand, J. et al. 2008, Nature, \textbf{454}, 735

\bibitem{dij04}
   Dijkstra, M., Haiman, Z., Rees, M.J, \& Weinberg, D.H. 
    2004, ApJ, \textbf{601}, 666

\bibitem{duc07}
   Duc, P.-A., \& Mirabel, I.F. 2007, A\&A, \textbf{333}, 813

\bibitem{fer94} 
   Ferguson, H.C., \& Binggeli, B. 1994, A\&ARev, \textbf{6}, 67

\bibitem{fum11} 
   Fumagalli, M., Gavazzi, G., Scaramella, R., \& Franzetti, P.
   2011, A\&A, \textbf{528}, 46

\bibitem{gar02}
   Garnett, D.R. 2002, ApJ, \textbf{581}, 1019

\bibitem{goe08}
   Goerdt, T., et al. 2008, \textbf{385}, 2136

\bibitem{got09}
   Gotthart,T., et al. 2009, AN, \textbf{330}, 1037

\bibitem{gre04}
   Grebel, E.K. \& Gallagher, J.S. 2004, ApJ, \textbf{610}, L89 

\bibitem{gre03}
   Grebel, E.K., Gallagher, J.S., \& Harbeck, D. 
   2003, AJ, \textbf{125}, 1966

\bibitem{gre09}
   Gregg, M.D., et al. 2009, AJ, \textbf{137}, 498

\bibitem{haa97} 
   Han, M., et al. 1997, AJ, \textbf{113}, 1001

\bibitem{har01}
   Harbeck, D., et al. 2001, AJ, \textbf{123}, 3092

\bibitem{har06}
   Harfst, S., Theis, C., \& Hensler, G. 
   2006, A\&A, \textbf{449}, 509

\bibitem{hen03} 
   Hensler, G. 2003, in: C. Charbonnel et al. (eds.), 
   ASP Conf. Ser. Vol., \textbf{304}, 371

\bibitem{hen98}
   Hensler, G., Dickow R., Junkes N., \& Gallagher, J.S. 
   1998, ApJ, \textbf{502}, L17                               

\bibitem{hens99} 
   Hensler, G., Rieschick A., \& Koeppen, J. 1999, 
   in: J.\ Beckman \& T.J.\ Mahoney (eds.), 
   {\it The Evolution of galaxies on Cosmological Timescales},
   ASP Conf. Ser., \textbf{187}, 214

\bibitem{hen04}
   Hensler, G., Theis, C., \& Gallagher, J.S. 
   2004, A\&A, \textbf{426}, 25 

\bibitem{hen10}
   Hensler, G. \& Recchi, S. 2010, 
   in:  K. Cunha, M. Spite \& B. Barbuy, (eds.), 
   {\it Chemical Abundances in the Universe: Connecting First Stars to Planets}, Proc. IAU Symp. No. \textbf{265}, p. 325

\bibitem{hun98}
   Hunter, D.A., et al. 1998, ApJ, \textbf{495}, L47

\bibitem{joh08}
  Johnston, K. et al.  2008, ApJ, \textbf{689}, 936

\bibitem{ken98} 
   Kennicutt, R.J. 1998, ApJ, \textbf{498}, 541

\bibitem{kho09}
   Khochfar, S. \& Silk, J. 2009, ApJ, \textbf{700}, L21

\bibitem{kob08}
   Kobulnicky, H.A. \& Skillman, E.D.  2008, AJ, \textbf{135}, 527

\bibitem{kob95}
   Kobulnicky, H.A., et al. 1995, AJ, \textbf{110}, 1116 

\bibitem{koc08}
   Koch, A., et al. 2008, AJ, \textbf{135}, 1580

\bibitem{koc09}
   Koch, A. 2009, Rev. Modern Astr., \textbf{18}, 675

\bibitem{koe05} 
   Koeppen, J. \& Hensler, G. 2005, A\&A, \textbf{434}, 531 

\bibitem{koe95}
   Koeppen, J., Theis, C., \& Hensler, G. 1995, A\&A, \textbf{296}, 99

\bibitem{kol09}
   Koleva, M., et al. 2009, MNRAS, \textbf{396}, 2133

\bibitem{kor85} Kormendy, J. 1985, ApJ, \textbf{295}, 73

\bibitem{kro10}
   Kroupa, P., et al. 2010, A\&A, \textbf{523}, A32

\bibitem{kue11}
   Kuehtreiber, M. 2011, Bachelor thesis, Univ. of Vienna

\bibitem{lm04}
   Lanfranchi, G.A., \& Matteucci, F. 2004, MNRAS, \textbf{351}, 1338

\bibitem{lar74} 
   Larson, R.B. 1974, MNRAS, \textbf{169},  229

\bibitem{lis06}
  Lisker, T., Grebel, E., \& Binggeli, B. 2006, AJ, \textbf{132}, 497

\bibitem{lis07} 
   Lisker, T., Grebel, E.K., Binggeli, B., \& Glatt, K.  
   2007, ApJ, \textbf{660}, 1186

\bibitem{lis09}
   Lisker, T., Janz, J., Hensler, G. et al. 2009, ApJ, \textbf{706}, 124

\bibitem{mlf99}
   MacLow, M.-M. \& Ferrara, A. 1999, ApJ, \textbf{513}, 142 

\bibitem{mar03}
   Marcolini, A., Brighenti, F., \& D'Ercole, A. 2003, MNRAS, \textbf{345}, 1329

\bibitem{mar08}
   Marcolini, A., et al. 2008, MNRAS, \textbf{386}, 2173

\bibitem{mar95} 
   Marlowe, A.T., Heckman, T.M., Wyse, R.F.G., \& Schommer R. 
    1995, ApJ, \textbf{438}, 563

\bibitem{mar02}
   Martin, C.L., Kobulnicky, H.A., \& Heckman, T.M. 
   2002, ApJ, \textbf{574}, 663 

\bibitem{mat98} 
   Mateo, M. 1998, ARA\&A, \textbf{36}, 435

\bibitem{may07}
   Mayer, L. et al. 2007, Nature, \textbf{445}, 738

\bibitem{met09}
   Metz, M., Kroupa, P., Theis, C., Hensler, G., \& Jerjen, H. 
    2009, ApJ, \textbf{697}, 269

\bibitem{meu98}
   Meurer, G.R., Staveley-Smith,L., \&  Killeen, N.E.B.  
   1998, MNRAS, \textbf{300}, 705

\bibitem{mic04} 
   Michielsen, D., et al. 2004, MNRAS, \textbf{353}, 1293 

\bibitem{mom05}
   Momany, Y., et al. 2005, A\&A, \textbf{439}, 111

\bibitem{mor00}
   Mori, M. \& Burkert, A. 2000, ApJ, \textbf{538}, 559

\bibitem{mue05}
    M\"uhle, S., Klein, U.,Wilcots, E. M., \& H\"uttermeister, S. 
    2005, AJ, \textbf{130}, 524

\bibitem{noe00}
   Noeske, K., et al. 2000, A\&A, \textbf{361}, 33

\bibitem{noe07}
   Noeske, K.G., et al. 2007, ApJ, \textbf{660}, L47

\bibitem{pag10} 
   Pagel, B.E.J. 2010, {\it Nucleosynthesis and galactic chemical evolution}, (Cambridge Univ. Press)

\bibitem{par10}
   Partl, A.M., Dall'Aglio, A., M\"uller, V., \& Hensler, G. 
   2010, A\&A, \textbf{524}, A85

\bibitem{pee08}
   Peeples, M.S., Pogge, R.W., \& Stanek, K.Z. 
   2008, ApJ, \textbf{685}, 904

\bibitem{pet93} 
   Petersen, R.C., \& Caldwell, N. 1993, AJ, \textbf{105}, 1411

\bibitem{ph11}
   Petrov, M. \& Hensler, G. 2012, ApJ, submitted

\bibitem{phi01}
   Phillipps, S., et al. 2001, ApJ, \textbf{560}, 201

\bibitem{pra08a}
  Prantzos, N. 2008a, in: C. Charbonnel \& J.-P. Zahn (eds.), 
  {\it Stellar Nucleosynthesis: 50 years after B$^2$FH; 
   EAS Publ. Ser.}, 32, 311

\bibitem{pra08b}
   Prantzos, N. 2008b, A\&A, \textbf{489}, 525

\bibitem{rea06}
   Read, J.I., et al. 2006, MNRAS, \textbf{366}, 429

\bibitem{rec06a} 
   Recchi, S., \& Hensler, G. 2006, A\&A, \textbf{445}, L39

\bibitem{rec07a}
   Recchi, S., \& Hensler, G. 2007, A\&A, \textbf{476}, 841

\bibitem{rec06b} 
   Recchi, S., Hensler, G., Angeretti, L., \& Matteucci, F. 
   2006, A\&A, \textbf{445}, 875

\bibitem{rec07b}
   Recchi, S., Theis, C., Kroupa, P., \& Hensler, G. 
   2007, A\&A, \textbf{470}, L5

\bibitem{rec09} 
   Recchi, S., Hensler, G., \& Anelli, D. 2009, arXiv:0901.1976

\bibitem{rev09}
   Revaz, Y., et al. 2009, A\&A, \textbf{501}, 189

\bibitem{roe05}
   Roediger, E., \& Hensler, G. 2005, A\&A, \textbf{431}, 85

\bibitem{sal08}
   Salvadori, S., Ferrara, A., \& Schneider, R. 
   2008, MNRAS, \textbf{386}, 348

\bibitem{sb84}
   Sandage, A. \& Binggeli, B. 1984, AJ, \textbf{89}, 919

\bibitem{sca06}
   Scannapieco, C., et al. 2006, MNRAS, \textbf{371}, 1125

\bibitem{sca10}
   Scannapieco, E., \& Brueggen, M. 2010, MNRAS, \textbf{405}, 1635

\bibitem{sea08}
   Seabroke, G.M., et al. 2008, MNRAS, \textbf{384}, 11

\bibitem{sti02}
   Stil, J. M., \& Israel, F. P. 2002, A\&A, \textbf{392}, 473

\bibitem{spo09}
   Spolaor, M., Proctor, R.N., Forbes, D.A., \&  Couch, W.J. 
   2009, ApJ, \textbf{691}, L138

\bibitem{str04}
   Strickland, D.K., et al. 2004, ApJ, \textbf{606}, 829

\bibitem{tolo09}
   Toloba, E., et al. 2009, ApJ, \textbf{707}, L17

\bibitem{tol11}
   Toloba, E., et al. 2011, A\&A, \textbf{526}, A114

\bibitem{tol09}
   Tolstoy, E., et al.  2009, ARA\&A, \textbf{47}, 371

\bibitem{tre04}
   Tremonti, C.A., et al. 2004, ApJ, \textbf{613}, 898

\bibitem{tre02} 
   Trentham, N. \& Tully, R.B. 2002, MNRAS, \textbf{335}, 712

\bibitem{tul08} 
   Tully, R.B. \& Trentham, N. 2008, AJ, \textbf{135}, 1488

\bibitem{zee01}
   van Zee, L. 2001, AJ, \textbf{121}, 2003

\bibitem{zee98a}
   van Zee, L., Salzer, J.J., \& Haynes M.P. 1998, ApJ, \textbf{497}, L1

\bibitem{zee98b}
   van Zee, L., Skillman, E.D., \& Salzer, J.J. 
   1998, AJ, \textbf{116}, 1186

\bibitem{zee98c}
   van Zee, L., Westphal, D., Haynes M.P., \& Salzer, J.J. 
   1998, AJ, \textbf{115}, 1000

\bibitem{zee04} 
   van Zee, L., Skillman, E.D., \& Haynes, M.P. 
   2004, ApJ, \textbf{218}, 211

\end{thebibliography}
